\documentclass[preprint,proceedings]{rmaa}

\suppressfulladdresses 

\usepackage{paralist}

\usepackage{psfrag,color}

\SetYear{}
\SetConfTitle{}

\title{Hydrogen neutral outflowing disks of B[e] supergiants} 

\author{
  M. Kraus,\altaffilmark{1} 
  M. Borges Fernandes,\altaffilmark{2}
  and F. X. de Ara\'ujo\altaffilmark{2}}

\altaffiltext{1}{Astronomical Institute AV \v{C}R, Fri\v{c}ova 298, 251\,65 
                 Ond\v{r}ejov, Czech Republic (kraus@sunstel.asu.cas.cz).}

\altaffiltext{2}{Observat\'orio Nacional-MCT, Rua General Jos\'e Cristino 77, 20921-400, S\~ao Cristov\~ao, Rio de Janeiro, Brazil (borges@on.br, araujo@on.br).}

\shortauthor{Kraus, Borges Fernandes, \& de Ara\'ujo}
\shorttitle{Hydrogen neutral outflowing disks}

\listofauthors{M. Kraus, M. Borges Fernandes, \& F. X. de Ara\'ujo}
\indexauthor{Kraus, M.}
\indexauthor{Borges Fernandes, M.}
\indexauthor{de Ara\'ujo, F. X.}

\abstract{The [O{\sc i}] line emission of the LMC B[e] supergiant R\,126 is 
modeled with an outflowing disk scenario. We find that hydrogen in the disk 
must be ionized by less than 0.1\%, meaning that the disk material is 
predominantly neutral. The free-free emission is calculated from the polar 
wind, and the minimum density contrast between disk and polar wind is found 
to be $\sim 10$.}

\resumen{}

\addkeyword{Stars: Mass loss}
\addkeyword{Stars: Winds, outflows}
\addkeyword{Supergiants}

\begin{document}
\maketitle

\section{Introduction \& Motivation}
\label{sec:intro}
The nature of the B[e] supergiant stars' disks is a long-standing problem (see
e.g. Kraus \& Miroshnichenko 2006). Recently, Porter (2003) investigated the 
possibility of dust formation in the pole-on seen disk of the LMC B[e] 
supergiant R\,126 for two different approaches: an
outflowing disk-forming wind, and a Keplerian viscous disk. He found that both
models failed in reproducing the observed dust and free-free emission
self-consistently. And he suggested a way out of this by allowing for
substantial alteration especially of the exponent of the radial density
distribution from a classical outflow with $\rho\sim r^{-2}$ to a considerably 
flatter one of $r^{-1.7}$. Such a flatter density profile guarantees that 
at every location in the disk a higher density (and hence opacity) is 
maintained, allowing for more efficient dust condensation and therefore for an 
enhanced dust emission over the disk. Such a modification finally resulted in a 
good fit to the SED of R\,126.

We have taken high- and low-resolution spectra of the same object using the 
FEROS and Boller \& Chivens spectrographs with a 2\arcsec~aperture diameter,
in order to observe the inner stellar wind material. These spectra show 
astonishingly strong [O{\sc i}] emission. Due to the nearly equal ionization
potentials of O and H, this emission must arise in regions where the wind 
material is neutral in hydrogen, i.e. in the disk. 
To test the modified outflowing disk scenario of Porter (2003), we took the
parameters of his best-fit model and calculated the emerging [O{\sc i}] line
luminosities. These turned out to be at least {\it a factor of 50 lower} than
the observed values, meaning that we need a much higher density within
the [O{\sc i}] line forming region than can be provided by Porter's disk model.

The most severe limitation in his model calculations was the assumption
that, as in the case of classical Be stars, the free-free emission arises in
the high density disk, while contributions from the polar wind are negligible.
Fitting the near-IR part of the SED with free-free emission from the outflowing
disk therefore determines the disk mass loss rate. Consequently, this value is
an {\it upper limit} for the disk density and hence opacity, and efficient dust 
formation can severely be hampered.


\section{The hydrogen neutral disk}
\label{sec:neutral}
In our outflowing disk model the density distribution follows from the mass
continuity equation, i.e. $\rho \sim r^{-2}$. The proportionality factor is 
not limited by the free-free emission, but follows purely from the fitting of 
the [O{\sc i}] line luminosities. The forbidden lines are optically thin. They
arise from the five lowest levels in O{\sc i} that are populated by collisions 
with free electrons. We calculate the level populations by solving the 
statistical equilibrium equations in a 5-level atom. Collision parameters are 
taken from Mendoza (1983) and atomic parameters from Wiese et al. (1966) and 
from Kafatos \& Lynch (1980). The oxygen and electron densities are 
parametrized in terms of the hydrogen particle density, i.e. $n_{\rm O}(r) = 
q_{\rm O} n_{\rm H}(r)$ and $n_{\rm e}(r) = q_{\rm e} n_{\rm H}(r)$. For 
$q_{\rm O}$ we use a typical LMC abundance value of 1/3 solar, with a solar 
oxygen abundance of $q_{\rm O, solar} = 6.76\times 10^{-4}$ (Grevesse \& Sauval 
1998), while $q_{\rm e}$ follows from the simultaneous fitting of the line 
ratios and luminosities of the [O{\sc i}] lines (see Kraus et al. 2007). 
The electron temperature, density parameter (i.e. mass flux over terminal 
velocity), and ionization fraction found for the disk-forming wind are given in 
Table~\ref{tab:param}.

\begin{table}[!t]\centering
  \setlength{\tabnotewidth}{\columnwidth}
  \tablecols{4}
  \caption{Disk and polar wind parameters.} \label{tab:param}
  \begin{tabular}{lccc}
    \toprule
    Wind & $T_{\rm e}$ & $F_{\rm m}/v_{\infty}$ & $q_{\rm e}$ \\
         & [K] & [g\,cm$^{-3}$] &  \\
    \midrule
     Polar        & 10\,000 & $2.0\times 10^{-12}$ & 1.0 \\
     Disk-forming & 8\,000  & $2.2\times 10^{-11}$ & $4.0\times 10^{-4}$ \\
    \bottomrule
  \end{tabular}
\end{table}

To compare our results with those of Porter (2003), we need to know the disk
mass loss rate. For this, the terminal velocity of the disk-forming wind 
needs to be known. The disk of R\,126 is seen pole-on and for the half opening 
angle we use Porter's value of $10^{\circ}$. Then we can estimate the outflow 
velocity from fitting the line profiles of the [O{\sc i}] lines (see 
Kraus et al. 2007). A maximum line-of-sight component of $\sim 2$\,km\,s$^{-1}$ 
might be hidden in the line profiles resulting in a disk terminal velocity
of about 11.5\,km\,s$^{-1}$. The total disk mass loss rate, which must be 
considered as a lower limit, results
in $\dot{M}_{\rm disk} \simeq 2.5\times 10^{-5}$\,M$_{\odot}$yr$^{-1}$. This
is a about a factor of 10 higher than the value used by Porter (2003).

Based on recent observations with the {\it Spitzer Space Telescope} Infrared
Spectrograph, Kastner et al. (2006) derived a total dust mass of the
disk around R\,126 of $\sim 3\times 10^{-3}$\,M$_{\odot}$. Converting their
dust mass into a disk mass loss rate we find $\dot{M}_{\rm disk}\simeq 3.4\times
10^{-4}$\,M$_{\odot}$yr$^{-1}$. This is even 10 times higher than our value,
confirming the correctness of our derived lower limit.




\section{The polar wind}
\label{sec:pol}
With a disk ionization fraction of less than 0.1\% the disk cannot be the main
source of the free-free emission. This emission must arise from the polar wind.
We fitted the SED of R\,126 in the optical and IR part (Kraus et al. 2007). 
The best fit polar wind parameters are summarized in Table~\ref{tab:param}. 
Comparing the density parameters in the disk and polar wind results in a 
density contrast of $\sim 10$. This is found to be a lower limit.

\section{Conclusions}
\label{sec:concl}
We proposed and tested the scenario of a hydrogen neutral disk for the LMC B[e]
supergiant R\,126 by modeling the line luminosities and line ratios of the
[O{\sc i}] emission lines resolved in our high-resolution optical spectra.
The parameters derived for the disk and wind of R\,126 are the following:
\begin{itemize}
\item We found that the [O{\sc i}] 6300\,\AA/5577\,\AA~line ratio is
very sensitive to the ionization fraction in the disk (see Kraus et al. 2007). 
From fitting the observed line ratio we can conclude that hydrogen in the disk 
is ionized by less than 0.1\%. This confirms that the disk is indeed 
predominantly neutral in hydrogen.
\item The disk mass loss rate of $\dot{M}_{\rm disk} \ga 2.5\times
10^{-5}$\,M$_{\odot}$yr$^{-1}$ found from our fitting is about a factor
of 10 higher than the value used by Porter (2003). This is in good
agreement with the postulated need for a higher disk density in order to fit
the [O{\sc i}] lines, and also in good agreement with the total dust
mass of about $3\times 10^{-3}$\,M$_{\odot}$ as derived by Kastner et al.
(2006). A disk with (much) higher density provides a much
better environment for efficient dust condensation than the disk model used
by Porter (2003).
\item The near-IR excess is fitted with free-free and free-bound emission
from the B-type line-driven polar wind. The resulting density contrast between
equatorial and polar wind is on the order of 10. This value is found to be
a lower limit.
\end{itemize}
To summarize, based on a detailed investigation of the emerging [O{\sc i}]
emission lines we found that the disk around the B[e] supergiant R\,126 is
neutral in hydrogen right from the stellar surface. Since all B[e] supergiants
show strong [O{\sc i}] line emission, we postulate that the hydrogen neutral
ouflowing disk scenario might also hold for the other members of the B[e]
supergiant class.

\begin{acknowledgements}
                                                                                
M.K. acknowledges financial support from GA \v{C}R grant number 205/04/1267, 
GA AV grant number KJB300030701, as well as from the LOC.
                                                                                
\end{acknowledgements}

\end{document}